\documentclass[aps,prd,12point,twocolumn,nofootinbib,showpacs,superscriptaddress]{revtex4-1}

\usepackage{graphicx}% Include figure files
\usepackage{dcolumn}% Align table columns on decimal point
\usepackage{tabu}
\usepackage{amsmath,amssymb}
\usepackage{float}
\usepackage{natbib}
\usepackage{balance}
\usepackage{mathrsfs}
\usepackage{bm}% bold math
\usepackage{lipsum}
\usepackage[dvipsnames]{xcolor}
\usepackage[title]{appendix}
\usepackage[breaklinks=true,pdfborder={0 0 0},bookmarksopen]{hyperref}
\hypersetup{
	colorlinks   = true, %Colours links instead of ugly boxes
	urlcolor     = blue, %Colour for external hyperlinks
	linkcolor    = blue, %Colour of internal links
	citecolor   = red %Colour of citations
}
\begin{document}

\title{More on spacetime thermodynamics in the light of Weyl transformations}

%\title{Jacobson found Einstein with the help of Clausius, Unruh and Bekenstein. What would Weyl, watching them, say?}

\author{Fay\c{c}al Hammad} \email{fhammad@ubishops.ca} 
\affiliation{Department of Physics and Astronomy \& STAR Research Cluster, Bishop's University, 2600 College Street, Sherbrooke, QC, J1M~1Z7 Canada} 
\affiliation{Physics Department, Champlain 
College-Lennoxville, 2580 College Street, Sherbrooke,  
QC, J1M~0C8 Canada}
\affiliation{D\'epartement de Physique, Universit\'e de Montr\'eal,\\
2900 Boulevard \'Edouard-Montpetit,
Montr\'eal, QC, H3T 1J4
Canada}

\author{Daniel Dijamco} \email{ddijamco18@ubishops.ca} 
\affiliation{Department of Physics and Astronomy \& STAR Research Cluster, Bishop's University, 2600 College Street, Sherbrooke, QC, J1M~1Z7 Canada}

\begin{abstract}
It was recently found that, after performing a Weyl conformal transformation, the familiar analogy between black hole mechanics and black hole thermodynamics becomes ambiguous. It was argued that this fact can be traced back to the fundamental dichotomy between matter and geometry, which is at the heart of Einstein's field equations. As a further study of this issue, we investigate here the general link between spacetime thermodynamics and Weyl transformations from two other angles. We first examine the conformal behaviour of the horizon entropy within Wald's approach based on the fundamental diffeomorphism symmetry of {\it pure} geometry. We then revisit --- using Weyl transformations --- Jacobson's derivation of Einstein's field equations, the starting point of which is precisely built on the fundamental dichotomy between matter and geometry. As a result, we show that in order for Jacobson's approach to be able to yield the right Einstein field equations in the conformal frame, a specific conformal behavior of the horizon temperature and its entropy --- different from what Wald's approach implies --- is required. The two approaches to spacetime thermodynamics become thus incompatible in the conformal frame. In addition, we show --- in greater detail in the conformal frame --- that in the presence of a null dust the thermodynamics approach for extracting Einstein's field equations necessarily fails. An extensive discussion of the whole issue is given. \end{abstract}

%PACS, the Physics and Astronomy Classification Scheme.
\maketitle
\section{Introduction}\label{SecI}
Among the various major outcomes of the original research on black hole thermodynamics \cite{BCH,Hawking,Bekenstein1,Bekenstein2, Israel} are the interpretation of black hole entropy as a Noether charge \cite{WaldEntropy,IyerWald} and the extraction of Einstein field equations for gravity from the second law of thermodynamics \cite{JacobsonPRL}. The import of these two outcomes is much appreciated when they are viewed as fundamental links between the dynamical equations describing spacetime, {\it i.e.}, gravitational physics, and the fundamental discipline of thermodynamics. The former is concerned with the {\it relation} between matter and the arena it gives rise to by its mere existence, while the latter is concerned with {\it pure} matter itself. 

The Noether charge of spacetime arises from the diffeomorphism symmetry of the latter. It was shown in Ref.~\cite{WaldEntropy} that in the presence of a black hole's horizon, such a charge represents the black hole's entropy. One (simple) way to look at this result is as follows. The spacetime diffeomorphism has always been there and has nothing to do with the presence of the black hole. However, as soon as a horizon, {\it i.e.}, a boundary, comes into being, the diffeomorphism symmetry suddenly manifests itself physically and becomes encoded on the horizon in the form of entropy. From this point of view, it becomes totally unexpected and remarkable that a symmetry, which is inherent in the arena on which matter exists, happens to encode an entropy --- which would represent a measure of ``disorder'' --- of such an arena. 

The link between spacetime curvature and matter is encapsulated in Einstein field equations. These dictate how geometry reacts to the presence of matter. In Ref.~\cite{JacobsonPRL}, it was shown that these equations can be extracted from the thermodynamic link --- more specifically, the Clausius relation --- between the heat transported by matter, on one hand, and the temperature and entropy associated to the geometry of the spacetime containing such matter, on the other. Now, at first sight this seems to contradict what we just saw in the previous paragraph. Indeed, if entropy is a manifestation of spacetime diffeomorphism only in the presence of a black hole horizon, how can one associate entropy to spacetime in the first place before the formation of a black hole? A possible way out is to recall that the Noether charge associated to diffeomorphism symmetry also manifests itself in the energy-momentum tensor of matter whenever there is some around. As it happens, in Jacobson's approach one indeed assumes a flow of matter (the fundamental matter-geometry dichotomy of Einstein equations is thus already assumed) along which the heat $\delta Q$ in Clausius' relation is carried through a local causal horizon $\mathscr{H}$. From this point of view, then, it is clear that Wald's and Jacobson's results cannot be derived from each other, but are rather complementary to each other. As such, one expects that by rescaling the spacetime metric the two results would still go hand in hand or, at least, would suggest a physical interpretation for extra terms, if any, as is the case of the conformally transformed Einstein equations. As we shall see, however, this is not the case.

Now, although such incompatibility can be expected given the different extents to which the two approaches rely on matter and geometry, one would still like to investigate the fate of their complementarity as well as their individual integrity in the conformal frame. As we shall see, both their complementarity and some of their integrity break down in the conformal frame.  

Our aim in this paper is therefore not to investigate the physical (non)equivalence of conformal frames (see e.g., Ref.~\cite{ConformalIssue} for a discussion on this issue), but is rather similar in spirit to Refs.~\cite{Hammad1, Hammad2, Hammad3, AugustPaper, OctoberPaper}. The best way to reveal the true nature of any concept in spacetime physics --- or, in this case, any incompatibility between two approaches towards understanding spacetime physics --- is in fact to view the concept, or apply the approaches, in a deformed setting that is capable of ``filtering out'' purely geometric from purely material entities. Such a deformed setting is, in fact, best accomplished by applying Weyl's conformal transformation, or ``mapping''. These transformations leave spacetime coordinates {\it intact} but change a spacetime metric $g_{ab}$ into a new metric $\tilde{g}_{ab}=\Omega^2g_{ab}$, where the conformal factor $\Omega(x)$ is a spacetime-dependent, non-vanishing, and everywhere regular function. Indeed, such an approach has successfully been applied to probe the true nature of some of the well-known quasilocal definitions of energy in general relativity \cite{Hammad1, Hammad2}, to study the behavior of black holes and wormholes under conformal transformations \cite{Hammad3}, to investigate the requirement for violating the null energy condition by traversable wormholes under different settings \cite{OctoberPaper} and, most importantly, to provide a different point of view on black hole mechanics and thermodynamics \cite{AugustPaper}.

In fact, in Ref.~\cite{AugustPaper} it was shown that a Weyl conformal transformation renders the familiar analogy between black hole mechanics and thermodynamics ambiguous. It was emphasized there that this issue arises whenever one mixes purely geometric concepts with purely material ones. Indeed, it was found that the conformal behavior of black hole thermodynamics depends on whether one studies the system based on the thermodynamics of the infalling matter or one focuses solely on the mechanics of the black hole's geometry. It was found that the black hole's entropy would transform like area under conformal transformations if it is identified with the purely geometric surface area, whereas if one traces it back to the infalling matter it becomes conformally invariant. Similarly, it was found that the black hole's temperature transforms differently depending on whether one identifies it with the purely geometric surface gravity or one takes it as a property of the infalling matter. On the basis of this conflict, it was argued in Ref.~\cite{AugustPaper} that when talking about black hole thermodynamics it would be more accurate to distinguish between the black hole's ``material'' thermodynamics and the black hole's ``geometric'' thermodynamics, for conformal transformations do not allow mixing between the two.

The apparent conflict that arises between Wald's approach and the conformal mapping is the following. Based on the very nature of Weyl's conformal transformation, one might think that after performing such a transformation one would necessarily find a different entropy when using Wald's approach because the spacetime coordinates remain intact, {\it i.e.}, the diffeomorphism symmetry is preserved, whereas the area of the boundary on which such an entropy manifests itself is modified. According to this argument, however, one naturally expects that under a conformal transformation entropy would scale like area. As we shall see, this is not the case. The reason behind this unexpected outcome will be provided.

The conflict between Jacobson's approach on the one hand, and the conformal mapping and Wald's approach on the other, is the following. In Jacobson's approach one uses Clausius' thermodynamics relation,  $\mathcal{T}{\rm d}S=\delta Q$, where $\mathcal{T}$ is temperature, to recover  
Einstein's field equations $R_{ab}-\tfrac{1}{2}g_{ab}R+\Lambda g_{ab}=8\pi T_{ab}$\footnote{We set $G=1$ throughout the paper.}, where $\Lambda$ is the usual unknown cosmological constant. The first natural question to ask then is whether one is able to recover the right conformally transformed Einstein field equations if one adopts again Clausius' relation after conformally transforming the geometry of spacetime. In fact, it is well known that the transformed Einstein equations take the following specific form in the conformal frame:
%%%%%%%%%%%%%%%%%%%%%%%%%%%%--------------------%%%%%%%%%%%%%%%%%%%%%%%%%%%%%%%%%%%%--------------------%%%%%%%%%%%%%%%%%%%%%%%%%%%%%%%%%%%%%%%%%%%%%%%%%%%%%
\begin{align}\label{ConfEQ}
&\tilde{R}_{ab}-\tfrac{1}{2}\tilde{R}\tilde{g}_{ab}+\frac{\Lambda}{\Omega^2}\tilde{g}_{ab}=8\pi\Omega^2 \tilde{T}_{ab}-\tilde{T}_{ab}^\Omega.
\end{align}
%%%%%%%%%%%%%%%%%%%%%%%%%%%%--------------------%%%%%%%%%%%%%%%%%%%%%%%%%%%%%%%%%%%%--------------------%%%%%%%%%%%%%%%%%%%%%%%%%%%%%%%%%%%%%%%%%%%%%%%%%%%%%
The extra term $\tilde{T}_{ab}^{\Omega}$ is interpreted as an {\it induced} energy-momentum tensor that arises due to the ``work'' done in deforming spacetime, and it is given in terms of the first and second derivatives of the conformal factor $\Omega$ as follows (see, e.g., Ref.~\cite{ConformalReference}. We suppressed here the extra ($-$) sign in the first term of Eq.~(IV.19) in that reference.),
%%%%%%%%%%%%%%%%%%%%%%%%%%%%--------------------%%%%%%%%%%%%%%%%%%%%%%%%%%%%%%%%%%%%--------------------%%%%%%%%%%%%%%%%%%%%%%%%%%%%%%%%%%%%%%%%%%%%%%%%%%%%%
\begin{equation}\label{InducedEMT}
\tilde{T}_{ab}^\Omega=\frac{2\tilde{\nabla}_a\tilde{\nabla}_b\Omega}{\Omega}-\tilde{g}_{ab}\left(\frac{2\widetilde{\Box}\Omega}{\Omega}-\frac{3\tilde{\nabla}_c\Omega\tilde{\nabla}^c\Omega}{\Omega^2}\right).
\end{equation}
%%%%%%%%%%%%%%%%%%%%%%%%%%%%--------------------%%%%%%%%%%%%%%%%%%%%%%%%%%%%%%%%%%%%--------------------%%%%%%%%%%%%%%%%%%%%%%%%%%%%%%%%%%%%%%%%%%%%%%%%%%%%%
As we shall see, {\it naively} adopting Clausius' relation in the conformal frame --- in the hope of preserving this nice thermodynamic origin for Einstein field equations --- does not lead to the right field equations.

The second natural question that imposes itself then is whether one is able to recover the right conformally transformed Einstein field equations if one accepts the use of whatever form is required of the Clausius relation in the conformal frame as long as it leads to the right field equations. As we shall see, the answer is yes in this case. However, the required conformal transformation of entropy that leads to the right equations is incompatible with what Wald's approach implies. Moreover, an additional issue arises in the presence of a null dust, a case for which the approach, by its very nature, is incapable of handling. A detailed discussion of all these issues will be provided.

The remainder of the paper is organized as follows. In Sec.~\ref{sec:II}, we briefly introduce Wald's approach for getting entropy from Noether's charge and then we adapt this method to the conformal frame. In Sec.~\ref{sec:III}, we first recall Jacobson's original extraction of Einstein field equations from Clausius' relation and then we investigate how the approach is affected by conformal transformations. We devote Sec.~\ref{sec:IV} to summarize and discuss all the issues exposed in the previous sections. 
%%%%%%%%%%%%%%%%%%%%%%%%%%%%%%%%%%%%%%%%%%%%%%%%%%%%%%%%%%%%%%%%%%%%%%%%%%%%%%%%%%%%%%%%%%%%%%%%%%%%%%%%%%%%%%%%%%%%%%%%%%%%%%%%%%%%%%%%%%%%%%%%%%%%%%%%%%%%%%%%%%%%%%%%%%%%%%%%%%%%%%%%%%%%%%%%%%%%%%%%%%%%%%%%%%%%%%%%%%%%%%%%%%%%%%%%%%%%%%%%%%%%%%%%%%%%%%%%%%%%%%%%%%%%%%%%%%%%%%%%%%%%%%%%
\section{Noether charge and Weyl rescaling}\label{sec:II}
Wald's method for computing entropy is based on the Noether charge associated with the diffeomorphism symmetry of spacetime. The purely geometric character of the approach makes it thus suitable for finding the entropy within any coordinate-independent theory of gravity \cite{WaldEntropy,IyerWald}. In other words, the approach works for any gravitational Lagrangian that is invariant under any change of coordinates. The approach even allows for a derivation of one class of modified gravity theories \cite{Fay}. In this paper, however, we are going to restrict ourselves to the Einstein-Hilbert Lagrangian (in four dimensions), {\it i.e.}, to general relativity (GR), as our purpose here is to learn more about spacetime as described by GR rather than to merely generalize a formalism which happens to work well in GR.

Wald's approach is best expressed and applied using differential forms. The gravitational Lagrangian $\mathcal{L}$ is thus viewed as a 4-form ${\bf L}=\mathcal{L}\,\boldsymbol{\epsilon}$ where $\boldsymbol{\epsilon}$ is the canonical volume form of spacetime. The variation of the Lagrangian under a diffeomorphism of infinitesimal generator $\xi^a$ then reads, $\delta{\bf L}={\bf E}\delta\phi+{\rm d}{\bf \Theta}$ \cite{WaldEntropy, IyerWald}. Here, the dynamical fields (the spacetime metric $g_{ab}$ and the matter fields $\psi$) of the theory are all included inside $\phi$. The equations of motion of the theory are then expressed by the vanishing of the 4-form ${\bf E}$. The 3-form ${\bf \Theta}(\phi,\pounds_\xi\phi)$ --- called the symplectic potential 3-form --- depends on the fields $\phi$ and their Lie derivative $\pounds_\xi\phi$ with respect to $\xi^a$. 

Next, a Noether 3-form current ${\bf J}$ can be associated to the generator $\xi^a$ and the fields $\phi$, and reads, ${\bf J}={\bf \Theta}(\phi,\pounds_\xi\phi)-\xi.{\bf L}$. One then is able to show that when the equations of motion of a diffeomorphism-invariant Lagrangian ${\bf L}$ are satisfied, i.e., ${\bf E}=0$, there exists a 2-form ${\bf Q}$ such that ${\bf J}={\rm d}{\bf Q}$ \cite{WaldEntropy, IyerWald}. The Noether charge associated to the diffeomorphism symmetry is thus identified with the 2-form $\bf Q$. 

Let us now apply this formalism to the Einstein-Hilbert Lagrangian and review how, in turn, entropy is extracted from the Noether charge ${\bf Q}$ \cite{WaldEntropy}. In this case, the Lagrangian reads in component form as,
%%%%%%%%%%%%%%%%%%%%%%%%%%%%--------------------%%%%%%%%%%%%%%%%%%%%%%%%%%%%%%%%%%%%--------------------%%%%%%%%%%%%%%%%%%%%%%%%%%%%%%%%%%%%%%%%%%%%%%%%%%%%%
\begin{equation}\label{HilbertLagrangian}
{\bf L}_{abcd}=\frac{\boldsymbol{\epsilon}_{abcd}}{16\pi}R,
\end{equation}
the variation of which yields the following symplectic potential 3-form \cite{IyerWald}:
%%%%%%%%%%%%%%%%%%%%%%%%%%%%--------------------%%%%%%%%%%%%%%%%%%%%%%%%%%%%%%%%%%%%--------------------%%%%%%%%%%%%%%%%%%%%%%%%%%%%%%%%%%%%%%%%%%%%%%%%%%%%%
\begin{equation}\label{SymplecticPotential}
{\bf \Theta}_{abc}=\frac{\boldsymbol{\epsilon}_{dabc}}{16\pi}\left(g^{de}\nabla^h\delta g_{eh}-g^{fh}\nabla^{d}\delta g_{fh}\right).
\end{equation}
Using this, one extracts, successively, the following Noether current 3-form and the Noether charge 2-form, respectively \cite{IyerWald},
%%%%%%%%%%%%%%%%%%%%%%%%%%%%--------------------%%%%%%%%%%%%%%%%%%%%%%%%%%%%%%%%%%%%--------------------%%%%%%%%%%%%%%%%%%%%%%%%%%%%%%%%%%%%%%%%%%%%%%%%%%%%%
\begin{align}\label{Current&Charge}
{\bf J}_{abc}&=\frac{\boldsymbol{\epsilon}_{dabc}}{8\pi}\,\nabla_e\left(\nabla^{[e}\xi^{d]}\right),\nonumber\\
{\bf Q}_{ab}&=-\frac{\boldsymbol{\epsilon}_{abcd}}{16\pi}\,\nabla^{c}\xi^{d}.
\end{align}
All one has to do now to find the entropy is to integrate this Noether charge 2-form $\bf Q$ over the bifurcation 2-surface $\Sigma$ of the horizon $\mathscr{H}$, on which the Killing vector $\xi^a$ vanishes. One finds \cite{WaldEntropy, IyerWald},
%%%%%%%%%%%%%%%%%%%%%%%%%%%%--------------------%%%%%%%%%%%%%%%%%%%%%%%%%%%%%%%%%%%%--------------------%%%%%%%%%%%%%%%%%%%%%%%%%%%%%%%%%%%%%%%%%%%%%%%%%%%%%
\begin{equation}\label{WaldEntropy}
\frac{\kappa}{2\pi} S=\int_{\Sigma}{\bf Q}=-\frac{\kappa}{16\pi}\int_{\Sigma}\boldsymbol{\epsilon}_{abcd}\,\boldsymbol{\epsilon}^{cd}=\frac{\kappa}{8\pi} A.
\end{equation}
%%%%%%%%%%%%%%%%%%%%%%%%%%%%--------------------%%%%%%%%%%%%%%%%%%%%%%%%%%%%%%%%%%%%--------------------%%%%%%%%%%%%%%%%%%%%%%%%%%%%%%%%%%%%%%%%%%%%%%%%%%%%%
To get the second equality one uses the familiar geometric identity that relates the Killing vector to the surface gravity, $\nabla^c\xi^d=\kappa\boldsymbol{\epsilon}^{cd}$, where $\boldsymbol{\epsilon}^{cd}$ is the bi-normal to $\Sigma$\footnote{It is clear, and important to note here, that extracting entropy according to the prescription (\ref{WaldEntropy}) does not in any way involve putting the surface gravity factor in by hand, as claimed in Ref.~\cite{BhattacharyaMajhi}. The surface gravity factor is already built-in inside the derivatives of the Killing vector $\xi^a$ and, hence, the Noether charge {\bf Q} already contains it also through its dependence on the derivatives of $\xi^a$. This observation allows us --- in contrast to what was argued in Ref.~\cite{BhattacharyaMajhi} --- to safely take Wald's approach as fundamental and apply it in the conformal frame where the Lagrangian has in fact the form of a scalar-tensor theory of gravity.} (see e.g., Ref.~\cite{Wald}). To get the last equality one uses that, thanks to the auxiliary null vector $N^a$ such that $\xi^aN_a=-1$, one has $\boldsymbol{\epsilon}^{cd}=\boldsymbol{\epsilon}^{cdef}N_e\xi_f$. Then, thanks to the identity $\boldsymbol{\epsilon}_{abcd}\boldsymbol{\epsilon}^{cdef}=-4\delta^{[e}_{a}\delta^{f]}_{b}$ \cite{Wald}, one is indeed left with $\int_\Sigma\boldsymbol{\epsilon}_{ab}=A$ from which the area $A$ of the horizon emerges. We see from this result that entropy should really be equal to $1/4$ of the area of the horizon.

Let us now adapt this method to the conformal frame. First, the conformal transformation $g_{ab}\rightarrow \tilde{g}_{ab}=\Omega^2g_{ab}$ transforms the Einstein-Hilbert Lagrangian (\ref{HilbertLagrangian}) into the following:
%%%%%%%%%%%%%%%%%%%%%%%%%%%%--------------------%%%%%%%%%%%%%%%%%%%%%%%%%%%%%%%%%%%%--------------------%%%%%%%%%%%%%%%%%%%%%%%%%%%%%%%%%%%%%%%%%%%%%%%%%%%%%
\begin{equation}\label{ConfHilbertLagrangian}
\tilde{{\bf L}}_{abcd}=\frac{\tilde{\boldsymbol{\epsilon}}_{abcd}}{16\pi}\left[\frac{\tilde{R}}{\Omega^2}+\frac{6}{\Omega^4}\tilde{\nabla}_a\Omega\tilde{\nabla}^a\Omega+6\tilde{\nabla}_a\left(\frac{\tilde{\nabla}^a\Omega}{\Omega^3}\right)\right].
\end{equation}
%%%%%%%%%%%%%%%%%%%%%%%%%%%%--------------------%%%%%%%%%%%%%%%%%%%%%%%%%%%%%%%%%%%%--------------------%%%%%%%%%%%%%%%%%%%%%%%%%%%%%%%%%%%%%%%%%%%%%%%%%%%%%
Notice that one usually discards the third term inside the square brackets because, being a total derivative, it constitutes a mere boundary contribution that has no effect in the search for the field equations. In Wald's approach, however, one needs to keep even a total derivative as it affects the Noether charge. Indeed, as shown in detail in Ref.~\cite{IyerWald}, by adding a total derivative to the Lagrangian, the latter transforms into $\tilde{\bf L}\rightarrow\tilde{\bf L}+{\rm d}\tilde{\boldsymbol{\mu}}$. The exact form ${\rm d}\tilde{\boldsymbol{\mu}}$ then also affects the Noether current in the same way, $\tilde{\bf J}\rightarrow \tilde{\bf J}+{\rm d}(\tilde{\xi}.\tilde{\boldsymbol \mu})$, which, in turn, leads to an extra term in the Noether charge of the form $\tilde{\bf Q}\rightarrow \tilde{\bf Q}+\tilde{\xi}.\tilde{\boldsymbol{\mu}}$. Notwithstanding the presence of this extra term, we can still, for our purposes here, ignore the third term in the Lagrangian (\ref{ConfHilbertLagrangian}). In fact, recall that in Wald's approach one extracts the horizon entropy from the Noether charge $\tilde{\bf Q}$ by integrating the latter over the bifurcation 2-surface $\tilde{\Sigma}$ of the horizon $\tilde{\mathscr{H}}$ on which the Killing vector $\tilde{\xi}^a$ vanishes. 

Thus, to simplify our expressions we are going to drop in what follows the last term in the Lagrangian (\ref{ConfHilbertLagrangian}). Performing then the usual variation on what is left of expression (\ref{ConfHilbertLagrangian}), by taking into account the fact that now we have two independent fields, i.e., the metric $\tilde{g}_{ab}$ and the scalar field $\Omega$, yields the following simple symplectic potential 3-form,
%%%%%%%%%%%%%%%%%%%%%%%%%%%%--------------------%%%%%%%%%%%%%%%%%%%%%%%%%%%%%%%%%%%%--------------------%%%%%%%%%%%%%%%%%%%%%%%%%%%%%%%%%%%%%%%%%%%%%%%%%%%%%
\begin{align}\label{ConfSymplecticPotential}
\tilde{{\bf \Theta}}_{abc}&=\frac{\tilde{\boldsymbol{\epsilon}}_{dabc}}{16\pi\Omega^4}\left[\tilde{g}^{de}\tilde{\nabla}^{h}\left(\Omega^2\delta \tilde{g}_{eh}\right)
-\tilde{g}^{fh}\tilde{\nabla}^{d}\left(\Omega^2\delta\tilde{g}_{fh}\right)\right]\nonumber\\
&\;\;\;+\frac{3\tilde{\boldsymbol{\epsilon}}_{dabc}}{4\pi\Omega^4}\,\tilde{\nabla}^d\Omega\,\delta\Omega.
\end{align}
%%%%%%%%%%%%%%%%%%%%%%%%%%%%--------------------%%%%%%%%%%%%%%%%%%%%%%%%%%%%%%%%%%%%--------------------%%%%%%%%%%%%%%%%%%%%%%%%%%%%%%%%%%%%%%%%%%%%%%%%%%%%%
Next, using this expression, one easily extracts the current 3-form, from which, in turn, one then extracts the Noether charge 2-form, respectively,
%%%%%%%%%%%%%%%%%%%%%%%%%%%%--------------------%%%%%%%%%%%%%%%%%%%%%%%%%%%%%%%%%%%%--------------------%%%%%%%%%%%%%%%%%%%%%%%%%%%%%%%%%%%%%%%%%%%%%%%%%%%%%
\begin{align}\label{ConfCurrent}
\tilde{{\bf J}}_{abc}&=\frac{\tilde{\boldsymbol{\epsilon}}_{dabc}}{8\pi}\,\tilde{\nabla}_e\left(\Omega^{-2}\tilde{\nabla}^{[e}\tilde{\xi}^{d]}+2\tilde{\xi}^{[e}\tilde{\nabla}^{d]}\Omega^{-2}\right),\nonumber\\
\tilde{{\bf Q}}_{ab}&=-\frac{\tilde{\boldsymbol{\epsilon}}_{abcd}}{16\pi\Omega^2}\left(\tilde{\nabla}^{c}\tilde{\xi}^{d}-\frac{4\,\tilde{\xi}^c\tilde{\nabla}^d\Omega}{\Omega}\right).
\end{align}
%%%%%%%%%%%%%%%%%%%%%%%%%%%%--------------------%%%%%%%%%%%%%%%%%%%%%%%%%%%%%%%%%%%%--------------------%%%%%%%%%%%%%%%%%%%%%%%%%%%%%%%%%%%%%%%%%%%%%%%%%%%%%

The final task now is to extract the entropy by integrating this Noether charge 2-form over the bifurcate 2-surface $\tilde{\Sigma}$ of the horizon $\tilde{\mathscr{H}}$. However, our calculation will be greatly simplified when using the very important fact that on the horizon we also have the identity, $\tilde{\xi}^{[a}\tilde{\nabla}^{b]}\Omega=0$. In fact, as shown in Ref.~\cite{AugustPaper}, this identity is the constraint one recovers from a conformal factor $\Omega$ that guarantees the existence of the Killing vector $\tilde{\xi}^a$ in the conformal frame. This identity is, in turn, what guarantees the uniformity of the black hole's temperature over its horizon in the conformal frame \cite{AugustPaper}. Thus, the black hole's entropy is finally given by,
%%%%%%%%%%%%%%%%%%%%%%%%%%%%--------------------%%%%%%%%%%%%%%%%%%%%%%%%%%%%%%%%%%%%--------------------%%%%%%%%%%%%%%%%%%%%%%%%%%%%%%%%%%%%%%%%%%%%%%%%%%%%%
\begin{equation}\label{ConfWaldEntropy}
\frac{\tilde{\kappa}}{2\pi} \tilde{S}=\int_{\tilde{\Sigma}}\tilde{{\bf Q}}=-\frac{\tilde{\kappa}}{16\pi\Omega^2}\int_{\tilde{\Sigma}}\tilde{\boldsymbol{\epsilon}}_{abcd}\,\tilde{\boldsymbol{\epsilon}}^{cd}=\frac{\tilde{\kappa}}{8\pi\Omega^2}\tilde{A}.
\end{equation}
%%%%%%%%%%%%%%%%%%%%%%%%%%%%--------------------%%%%%%%%%%%%%%%%%%%%%%%%%%%%%%%%%%%%--------------------%%%%%%%%%%%%%%%%%%%%%%%%%%%%%%%%%%%%%%%%%%%%%%%%%%%%%
In the second step we have used an exactly similar geometric identity that relates the Killing vector to the surface gravity as done below Eq.~(\ref{WaldEntropy}), but written here in the conformal frame: $\tilde{\nabla}^c\tilde{\xi}^d=\tilde{\kappa}\tilde{\boldsymbol{\epsilon}}^{cd}$. 

We clearly see that according to this result the entropy in the conformal frame is not equal to $1/4$ of the transformed area $\tilde{A}$ but is rather equal to $\tilde{A}/(4\Omega^2)$. Now this is exactly what has been found in Ref.~\cite{AugustPaper} by studying the conformal transformation of black hole mechanics. Furthermore, this result is actually identical to what has also been found in Refs.~\cite{BhattacharyaMajhi,JacobsonKangMyers,Kang,RaraoniNielsen} by following a different route at the price of introducing the scalar field $\phi=\Omega^{-2}$ and then treating the resulting Lagrangian (\ref{ConfHilbertLagrangian}) as a scalar-tensor theory of gravity. However, our approach has the benefit of making transparent the various contributions of the conformal factor to the derivation, first inside the charge and its integral, and then in the final result. It is this explicit manifestation at various stages of the derivation that allows us to appreciate the role of the conformal factor in shaping the black hole's thermodynamics in the conformal frame. 

Therefore, although this result seems at odds and contrary to what one expects to find in the conformal frame based on black hole thermodynamics' area law, one can always argue, as was done in Refs.~\cite{BhattacharyaMajhi,JacobsonKangMyers,Kang,RaraoniNielsen}, that this comes about because one is not in GR anymore but in its scalar-tensor extension. A far more physically convincing justification for this result, however, is that the latter actually just confirms, as explained in the second paragraph above Eq.~(\ref{ConfEQ}), the very nature of Wald's approach. In fact, a conformal transformation has no effect on the diffeomorphism symmetry of spacetime from which the Noether charge arises in which, in turn, entropy is encoded. Thus, although the area is modified after a conformal transformation, the symmetry of spacetime remains intact. This fact is indeed clearly manifested in our result since $\tilde{S}=\tilde{A}/(4\Omega^2)=A/4=S$. According to this, one then simply concludes that according to Wald's approach entropy is encoded in the symmetry of spacetime and is, hence, invariant under conformal transformations regardless of the fate of the area on which it originally manifested itself.

Now if it were only a matter of justifying the fate of spacetime entropy within Wald's approach, the above interpretation would be amply satisfactory. Unfortunately, a new issue arises as soon as one confronts the above result with the other approach to spacetime thermodynamics; namely, Jacobson's extraction of Einstein's field equations from a thermodynamic relation, to which we turn now.
%%%%%%%%%%%%%%%%%%%%%%%%%%%%%%%%%%%%%%%%%%%%%%%%%%%%%%%%%%%%%%%%%%%%%%%%%%%%%%%%%%%%%%%%%%%%%%%%%%%%%%%%%%%%%%%%%%%%%%%%%%%%%%%%%%%%%%%%%%%%%%%%%%%%%%%%%%%%%%%%%%%%%%%%%%%%%%%%%%%%%%%%%%%%%%%%%%%%%%%%%%%%%%%%%%%%%%%%%%%%%%%%%%%%%%%%%%%%%%%%%%%%%%%%%%%%%%%%%%%%%%%%%%%%%%%%%%%%%%%%%%%%%%%%
\section{``Einstein equation of state'' and Weyl rescaling}\label{sec:III}
Jacobson's approach for extracting Einstein field equations is based on Clausius' relation $\mathcal{T}{\rm d}S=\delta Q$ --- whence the name ``Einstein equation of state'' \cite{JacobsonPRL} --- applied on the in-falling matter across a Rindler horizon $\mathscr{H}$. One should then identify $\mathcal{T}$ with the Unruh temperature $\mathcal{T}=\kappa/2\pi$ associated to the in-falling observer, where $\kappa$ is the surface gravity associated with the Rindler horizon. As a prelude to the issues we are going to face in this section when trying to adapt the approach to the conformal frame, it is important to provide here the following few remarks. 

First, the Clausius relation is capable of giving rise to Einstein field equations only when one assigns to the left-hand side of the relation geometric properties while to its right-hand side one assigns material properties. This act, however, can only be justified by the already known fact that Einstein field equations are dualistic in nature and that they alone can introduce this mysterious bridge between matter and geometry. 

Second, assigning temperature to a geometric entity is possible only if one identifies $\mathcal{T}$ with Unruh's temperature $\kappa/2\pi$. For in this case, one evokes the surface gravity $\kappa$ which is indeed a geometric entity and has nothing to do with Einstein equations in the first place. Notwithstanding this purely geometric character of surface gravity, one should keep in mind that temperature acquires its real meaning only with reference to matter. Therefore, this term $\mathcal{T}$ on the left-hand side of the Clausius relation does indeed conceal an unjustified, but deep, relation between geometry and matter that is {\it a priori} independent of Einstein field equations. 

Finally, assigning entropy to geometry is possible only if one identifies $S$ with Bekenstein's entropy which is proportional to the area $A$ of the horizon. However, this relation is much more transparent in the case of a black hole horizon as the latter's surface area indeed increases by just the right amount to save the second law of thermodynamics. Unfortunately, as this fact relies on black holes, which are solutions of Einstein field equations, one quickly runs into a circular reasoning when relying on such a relation to derive Einstein field equations. A possible way to avoid such a circularity trap is therefore to evoke entanglement entropy. In fact, as entanglement entropy also yields an area law \cite{Sorkin,t'Hooft,Bombelli,Srednicki,Calabresse,Callan,Holzhey} (see the nice reviews \cite{Viola, Solodukhin} for more recent references), one could safely translate into geometry the second term ${\rm d}S$ on the left-hand side of the Clausius relation without invoking gravity in the first place. This point of view does, however, have its own limitations as well. The limitations arise this time because one justifies the entropy increase ${\rm d}S$ on the left-hand side of the Clausius relation by the heat flow $\delta Q$ brought in by the in-falling matter of the right-hand side. Thus, one could then ask: What would happen if one uses instead the usual thermodynamic entropy we have about the in-falling matter prior to its horizon crossing, an entropy which has nothing to do with entanglement? It must be mentioned here, however, that another derivation of Einstein equations based specifically on entanglement entropy has been given in Ref.~\cite{Jacobson2016}. Being interested here, as a first step, only in the relation between spacetime dynamics and classical thermodynamics, we are going to leave the issue of entanglement for a future work.

Let us just ignore for now all these issues and focus here instead on adopting the approach of Ref.~\cite{JacobsonPRL} to the conformal frame. In the original frame one thus starts from Clausius relation in the form $\mathcal{T}{\rm d}S=\delta Q$. Then, one has to identify the temperature $\mathcal{T}$ with $\kappa/2\pi$ and substitute $\int_\mathscr{H}T^{ab}\xi_a{\rm d}\Sigma_b$ for the change in the heat $\delta Q$. This is because the heat flow is carried across the Rindler horizon $\mathscr{H}$ by the energy-momentum tensor $T_{ab}$ of matter along the Killing vector $\xi^a$ which becomes null on the horizon, and ${\rm d}\Sigma_b$ is a 2-surface element on the horizon. From this, one then goes through the following seven steps to arrive at Einstein equations \cite{JacobsonPRL}:
%%%%%%%%%%%%%%%%%%%%%%%%%%%%--------------------%%%%%%%%%%%%%%%%%%%%%%%%%%%%%%%%%%%%--------------------%%%%%%%%%%%%%%%%%%%%%%%%%%%%%%%%%%%%%%%%%%%%%%%%%%%%%
\begin{align}\label{Steps}
&\;\frac{\kappa}{2\pi}{\rm d}S=\int_{\mathscr{H}} T^{ab}\xi_a{\rm d}\Sigma_{b}\nonumber\\
\Rightarrow&\;\frac{\kappa}{2\pi}\eta\,\delta A=-\kappa\int_{\mathscr{H}}\lambda T^{ab}k_ak_b{\rm d}\lambda{\rm d}A\nonumber\\
\Rightarrow&\;\frac{\eta}{2\pi}\int_{\mathscr{H}}\theta{\rm d}\lambda{\rm d}A =-\int_{\mathscr{H}}\lambda T_{ab}k^ak^b{\rm d}\lambda{\rm d}A\nonumber\\
\Rightarrow&\;\frac{\eta}{2\pi}\int_{\mathscr{H}}-\lambda R_{ab}k^ak^b{\rm d}\lambda{\rm d}A =-\int_{\mathscr{H}}\lambda T_{ab}k^ak^b{\rm d}\lambda{\rm d}A\nonumber\\
\Rightarrow&\;\frac{\eta}{2\pi}R_{ab}k^ak^b = T_{ab}k^ak^b\nonumber\\
\Rightarrow&\;R_{ab}+fg_{ab} = \frac{2\pi}{\eta}T_{ab}\nonumber\\
\Rightarrow&\;R_{ab}-\frac{1}{2}Rg_{ab} + \Lambda g_{ab}=\frac{2\pi}{\eta}T_{ab}.
\end{align}
%%%%%%%%%%%%%%%%%%%%%%%%%%%%--------------------%%%%%%%%%%%%%%%%%%%%%%%%%%%%%%%%%%%%--------------------%%%%%%%%%%%%%%%%%%%%%%%%%%%%%%%%%%%%%%%%%%%%%%%%%%%%%
In the second line, one uses on the left-hand side the area law for entropy, $S=\eta A$, with some unknown constant of proportionality $\eta$, while on the right-hand side one uses $\xi^a=-\kappa\lambda k^a$ as well as ${\rm d}\Sigma_a=k_a{\rm d}\lambda{\rm d}A$. This is because the Rindler horizon $\mathscr{H}$ has $k^a$ as the null vector generator and is parametrized by the affine parameter $\lambda$. In the third line, one uses the fact that a null congruence expansion $\theta$ represents the relative variation $\delta A/A$ of the area $A$ in the transverse space. Indeed, for an infinitesimal area ${\rm d}A$ the relative variation along the affine parameter $\lambda$ would then be given by $\theta=\tfrac{\rm d}{\rm d \lambda}({\rm d}A)/{\rm d}A$, which leads to that integral over $\mathscr{H}$ on the left-hand side. In the fourth line, one integrates Raychaudhuri's equation $\tfrac{\rm d}{\rm d\lambda}\theta=-\tfrac{1}{2}\theta^2-\sigma_{ab}\sigma^{ab}+\omega_{ab}\omega^{ab}-R_{ab}k^{a}k^{b}$ (see, e.g., Ref.~\cite{Poisson}) to find $\theta$ by neglecting the second-order terms $\theta^2$, $\sigma^2$ and $\omega^2$. In the fifth line one suppresses the integration and sets equal the integrands of both sides as the equation is supposed to hold for an arbitrary integration over $\lambda$ and $A$. In the sixth line, one simplifies the $k$'s by taking into account that $k^a$ is null and, therefore, one could have an extra arbitrary scalar $f$ which depends on the metric and its derivatives, and which appears on the left-hand side multiplied by $g_{ab}$. Finally, in the last line one determines the unknown function $f$ by using, on the one hand, the contracted Bianchi identity $\nabla^bR_{ab}=\tfrac{1}{2}\nabla_{a}R$ and, on the other, the conservation of energy-momentum, $\nabla^bT_{ab}=0$. The last line represents Einstein field equations provided one sets the unknown proportionality constant $\eta$ between entropy and area equal to $1/4$ \cite{JacobsonPRL}. 

Let us now examine what happens in the conformal frame. After a Weyl conformal transformation, the new frame acquires the metric $\tilde{g}_{ab}$ and we have the following relations between the old and the new quantities that should appear in the extraction of Einstein equations:
%%%%%%%%%%%%%%%%%%%%%%%%%%%%--------------------%%%%%%%%%%%%%%%%%%%%%%%%%%%%%%%%%%%%--------------------%%%%%%%%%%%%%%%%%%%%%%%%%%%%%%%%%%%%%%%%%%%%%%%%%%%%%
\begin{align}\label{ConfTransfos}
&\tilde{\xi}^a=\Omega^{-1}\xi^a,\quad \tilde{k}^a=\Omega^{-1}k^a,\quad\tilde{\lambda}=\Omega\lambda, \quad \tilde{A}=\Omega^2A,\nonumber\\
&\tilde{T}_{ab}=\Omega^{-2}T_{ab},\quad\tilde{\nabla}^{b}\tilde{T}_{ab}=-\tilde{T}\Omega^{-1}\tilde{\nabla}_a\Omega.
\end{align}
%%%%%%%%%%%%%%%%%%%%%%%%%%%%--------------------%%%%%%%%%%%%%%%%%%%%%%%%%%%%%%%%%%%%--------------------%%%%%%%%%%%%%%%%%%%%%%%%%%%%%%%%%%%%%%%%%%%%%%%%%%%%%
In addition, in order for the conformal factor $\Omega$ to guarantee the existence of a Killing vector $\tilde{\xi}^a$ in the conformal frame, such a conformal factor $\Omega$ must be chosen such that $\xi^a\nabla_a\Omega=0$, which implies that $\tilde{\xi}^a\tilde{\nabla}_a\Omega=0$ (and hence also $\tilde{k}^a\tilde{\nabla}_a\Omega=0$) must be satisfied \cite{AugustPaper}.

Let us now examine what Jacobson's approach yields depending on what behavior we assign to entropy and temperature under the conformal transformations. To begin with, we need to decide upon the form Clausius' relation should have in the conformal frame. Because the observer in the conformal frame is supposed to not notice any change in the physics, it is natural to expect that the Clausius relation reads in the conformal frame as follows:
\begin{equation}\label{NaiveClausius}
\tilde{\mathcal{T}}{\rm d}\tilde{S}=\delta\tilde{Q}.
\end{equation}
It is straightforward to see, however, that starting from such a relation and going through the seven steps (\ref{Steps}) above, all one has to do is put tildes on all the symbols at each step and end up with Einstein equations in the form, $\tilde{R}_{ab}-\tfrac{1}{2}\tilde{R}\tilde{g}_{ab}+\Lambda \tilde{g}_{ab}=8\pi \tilde{T}_{ab}$. This is obviously not what one should find as this is different from Eq.~(\ref{ConfEQ}). We conclude from this that the observer in the conformal frame {\it cannot} simply use the Clausius relation in his/her reference frame as the starting point to extract Einstein field equations. 

The other possibility then is to just start from Clausius' relation as it holds in the original frame, {\it i.e.}, in the form $\mathcal{T}{\rm d}S=\delta Q$, and then to let it have whatever form it should have in the conformal frame as a consequence of the various geometric identifications required by Jacobson's approach. This is actually what one usually does when going from Einstein field equations in a given frame to the field equations in the conformal frame. Indeed, one just starts out by assuming in the original frame an energy-momentum tensor $T_{ab}$ that does satisfy Einstein equations in their familiar form. Then, one expresses all the quantities on both sides of such equations in terms of the corresponding conformal frame quantities. The resulting equations take the form (\ref{ConfEQ}) and are considered to represent Einstein equations in the conformal frame regardless of one's prior expectations. Nevertheless, as mentioned in the Introduction, the extra ``unexpected'' terms on the left-hand side of such equations easily acquire a nice interpretation when moved to the right-hand side that contains matter. They are thought of as representing an induced energy-momentum tensor due to the work done in deforming the spacetime. Unlike Einstein equations, however, Clausius' relation, $\mathcal T{\rm d}S=\delta Q$, is already a material-to-material relation. In other words, both sides of this relation are concerned with properties of matter. As such, one obtains, after a conformal transformation, a different final equation depending on which conformal behavior one decides to assign to each material property inside the equation. Before we adopt this philosophy and expose its consequences, let us first quickly discuss the other plausible possibilities by considering each term of Clausius' relation individually.

As for the conformal behavior of the heat flow $\delta Q$, one does not have much choice as it involves an energy-momentum tensor, {\it i.e.}, matter, and a purely geometric hypersurface, the conformal behavior of which is already fixed by definition. For the temperature $\mathcal{T}$ and the entropy $S$, however, things are more flexible. The first possibility is to let both of them be conformally invariant, {\it i.e.}, $\tilde{\mathcal{T}}=\mathcal{T}$ and $\tilde{S}=S$. In this case, instead of starting our derivation from Clausius' relation $\mathcal{T}{\rm d}S=\delta Q$, we would just use the would-be equivalent form,
\begin{equation}\label{Clausius1}
\tilde{\mathcal{T}}{\rm d}\tilde{S}=\delta Q.
\end{equation}
The other possibility is that either temperature or entropy are invariant, but not both. That is, either one has $\tilde{\mathcal{T}}=\mathcal{T}$, or one has $\tilde{S}=S$. In this case, one would start with Clausius' relation in either one of the following would-be equivalent forms, respectively,
\begin{equation}\label{Clausius2&3}
\tilde{\mathcal{T}}{\rm d}S=\delta Q\quad {\rm or} \quad \mathcal{T}{\rm d}\tilde{S}=\delta Q.
\end{equation}
It turns out that none of these three forms of Clausius' relation allows to recover the right Einstein field equations in the conformal frame. Let us then not make any prior choice now, and just start with the usual form of the Clausius relation, $\mathcal{T}{\rm d}S=\delta Q$, which we know is valid in the original frame.

Thus, we are going to extract the field equations in the conformal frame by starting from the second line of Eqs.~(\ref{Steps}) and then express each of the factors on both sides of that line in terms of the corresponding conformal frame quantities using the relations (\ref{ConfTransfos}). For surface gravity we use the relation $\tilde{\kappa}=\Omega^{-1}\kappa$ \cite{AugustPaper}. This leads to,
%%%%%%%%%%%%%%%%%%%%%%%%%%%%--------------------%%%%%%%%%%%%%%%%%%%%%%%%%%%%%%%%%%%%--------------------%%%%%%%%%%%%%%%%%%%%%%%%%%%%%%%%%%%%%%%%%%%%%%%%%%%%%%
\begin{align}\label{ConfSteps}
&\;\frac{\tilde{\kappa}}{2\pi\Omega}\eta\,\delta \tilde{A}=-\Omega\tilde{\kappa}\int_{\mathscr{H}}\tilde{\lambda} \tilde{T}_{ab}\tilde{k}^a\tilde{k}^b{\rm d}\tilde{\lambda}{\rm d}\tilde{A}\nonumber\\
\Rightarrow&\;\frac{\eta}{2\pi}\int_{\mathscr{H}}\tilde{\theta}{\rm d}\tilde{\lambda}{\rm d}\tilde{A} =-\Omega^2\int_{\mathscr{H}}\tilde{\lambda} \tilde{T}_{ab}\tilde{k}^a\tilde{k}^b{\rm d}\tilde{\lambda}{\rm d}\tilde{A}\nonumber\\
\Rightarrow&\;\frac{\eta}{2\pi}\int_{\mathscr{H}}-\tilde{\lambda} \tilde{R}_{ab}\tilde{k}^a\tilde{k}^b{\rm d}\tilde{\lambda}{\rm d}\tilde{A} =-\Omega^2\int_{\mathscr{H}}\tilde{\lambda}\tilde{T}_{ab}\tilde{k}^a\tilde{k}^b{\rm d}\tilde{\lambda}{\rm d}\tilde{A}\nonumber\\
\Rightarrow&\;\frac{\eta}{2\pi}\tilde{R}_{ab}\tilde{k}^a\tilde{k}^b = \Omega^2\tilde{T}_{ab}\tilde{k}^a\tilde{k}^b\nonumber\\
\Rightarrow&\;\tilde{R}_{ab}+f\tilde{g}_{ab}+W_{ab} = \frac{2\pi}{\eta}\Omega^2\tilde{T}_{ab}\nonumber\\
\Rightarrow&\;\tilde{R}_{ab}-\frac{1}{2}\tilde{R}\tilde{g}_{ab}+\frac{\Lambda}{\Omega^2}\tilde{g}_{ab}+\tilde{T}^\Omega_{ab}=\frac{2\pi}{\eta}\Omega^2\tilde{T}_{ab}.
\end{align}
%%%%%%%%%%%%%%%%%%%%%%%%%%%%--------------------%%%%%%%%%%%%%%%%%%%%%%%%%%%%%%%%%%%%--------------------%%%%%%%%%%%%%%%%%%%%%%%%%%%%%%%%%%%%%%%%%%%%%%%%%%%%%
In the third line we have used Raychaudhuri's equation in the conformal frame \cite{OctoberPaper}. In the fifth line, we have introduced a scalar $f$ and a 2-tensor $W_{ab}$ that might both depend {\it a priori} on the metric and the conformal factor $\Omega$ and their derivatives. Furthermore, the 2-tensor must be such that $W_{ab}\tilde{k}^{a}\tilde{k}^{b}=0$. In the last line, we have taken into account the fact that the divergences of both sides of the whole line should agree. We have then solved the resulting condition to find what $f$ and this 2-tensor are. The detailed calculations are given in Appendix \ref{A}. We thus see that the induced energy-momentum tensor $\tilde{T}_{ab}^{\Omega}$ does indeed emerge, as it should, in agreement with the correct field equations (\ref{ConfEQ}).

From the first line of Eqs.~(\ref{ConfSteps}) it is clear what form of the Clausius relation should be adopted in the conformal frame of the observer for the latter to recover the right Einstein field equations. It should be of the specific form: \begin{equation}\label{RequiredClausius}
\tilde{\mathcal{T}}{\rm d}\tilde{S}=\Omega^2\delta\tilde{Q}
\end{equation} 
It is easy to check by following the successive steps (\ref{ConfSteps}) that any other form of the conformal behavior of temperature $\mathcal{T}$ and entropy $S$ would not allow one to recover the right conformal Einstein equations. 

Now, this specific form (\ref{RequiredClausius}) of the Clausius relation in the conformal frame simply does not make any sense when the Clausius relation is viewed from the familiar status of a matter-to-matter thermodynamic relation. In fact, based on the energy dimensions of heat and temperature, and the dimensionless nature of entropy, one would rather expect the latter three quantities to transform as, 
\begin{equation}\label{NaturalExpectations}
\tilde{\mathcal{T}}=\mathcal{T}/\Omega,\quad{\rm d}\tilde{S}={\rm d}S,\quad\delta\tilde{Q}=\delta Q/\Omega.
\end{equation}
However, this conformal behavior would straightforwardly lead to the form (\ref{NaiveClausius}) of Clausius' relation in the conformal frame, a form which we already saw does not lead to the right field equations in such a frame. 

Therefore, this result shows that not only both temperature and entropy should be affected by the conformal transformation, but that, in addition, these should be affected in a way that is at odds with what Wald's approach implies. For Jacobson's approach to work in the conformal frame, temperature and entropy can only scale according to their geometric representations in terms of surface gravity and area. That is, $\tilde{\mathcal{T}}=\tilde{\kappa}/2\pi=\mathcal{T}/\Omega$ and that ${\rm d}\tilde{S}=\eta\delta\tilde{A}=\Omega^2{\rm d}S$, which, together with $\delta\tilde{Q}=\delta Q/\Omega$, do indeed lead to the strange form (\ref{RequiredClausius}). It is no wonder, however, that this should be the case as one has to convert the familiar matter-to-matter Clausius' thermodynamic relation into a matter-to-geometry relation by hand in order to recover the peculiar matter-geometry duality which only Einstein equations could ``motivate''.

Another issue which arises besides the ones discussed above, and which is still related to this crucial matter-geometry dichotomy, is the following. One of the defining steps in Eqs.~(\ref{ConfSteps}) for the extraction of Einstein equations consists in simplifying the null vector $\tilde{k}^a$ from both sides of the fifth line. That step is actually not at all trivial. Suppose indeed that there exists some null dust, with energy-momentum tensor $T^{\rm null}_{ab}$, contained inside the total energy-momentum tensor of matter $T_{ab}$. Then, as shown in Appendix \ref{B}, the conformal frame Einstein equations one obtains are the incorrect equations:
\begin{equation}\label{NullDustIssue}
\tilde{R}_{ab}-\tfrac{1}{2}\tilde{g}_{ab}\tilde{R}+\frac{\Lambda}{\Omega^2}\tilde{g}_{ab}+\tilde{T}_{ab}^\Omega = \frac{2\pi}{\eta}\left(\Omega^2\tilde{T}_{ab}+c\,\tilde{T}^{\rm null}_{ab}\right),
\end{equation}
for any arbitrary constant $c$\footnote{This remark actually also applies to the original frame derivation in which one follows the steps (\ref{Steps}). In this case, the crucial step is the one made in the sixth line, and the resulting incorrect Einstein equations with a null dust are, $R_{ab}-\tfrac{1}{2}g_{ab}R+\Lambda = (2\pi/\eta)\left(T_{ab}+c\,T^{\rm null}_{ab}\right)$, for an arbitrary constant $c$.}. This result can be traced back to the fact that, because of the very nature of the approach which consists in identifying at some point material properties with geometry, one is necessarily confronted with the null quantities' ambiguity by trying to extract an identity in a reverse way from a link that only Einstein field equations could have built.
%%%%%%%%%%%%%%%%%%%%%%%%%%%%%%%%%%%%%%%%%%%%%%%%%%%%%%%%%%%%%%%%%%%%%%%%%%%%%%%%%%%%%%%%%%%%%%%%%%%%%%%%%%%%%%%%%%%%%%%%%%%%%%%%%%%%%%%%%%%%%%%%%%%%%%%%%%%%%%%%%%%%%%%%%%%%%%%%%%%%%%%%%%%%%%%%
\section{Summary \& discussion}\label{sec:IV}
Spacetime thermodynamics has here been examined from an as yet different angle. The fate of both black hole entropy, as it emerges from Wald's method based on the Noether charge associated with the diffeomorphism invariance of GR, and Jacobson's extraction of Einstein field equations from Clausius' thermodynamic relation have been investigated under Weyl conformal transformations. Our investigation corroborates what has been concluded in Ref.~\cite{AugustPaper}. In the latter, it was pointed out that one has to distinguish what one would call the black hole's ``material'' thermodynamics from what one should rather call the black hole's geometric thermodynamics.

Our investigation of the effect of the spacetime rescaling on Wald's entropy showed that such an entropy does not scale like an area under a conformal transformation, but rather remains invariant. Although this result clashes with the first law of black hole mechanics, it is completely expected based on the very nature of the approach. Indeed, we saw that Wald's approach brings to light the transmutation of the spacetime diffeomorphism symmetry into entropy and, hence, that the latter should, just like the former, be unaffected by Weyl transformations. This result actually acquires its full meaning {\it only} when compared to what was found in Ref.~\cite{AugustPaper}. In fact, it was found in the latter reference that, while within black hole mechanics entropy should transform as area, within black hole material thermodynamics entropy should actually be invariant. By recalling that diffeomorphism symmetry manifests itself in the existence of the conserved energy-momentum tensor of matter the piece of the puzzle suddenly takes its place as it becomes then clear that Wald's approach is actually equivalent to the material thermodynamics. This comes about because, in contrast to black hole mechanics, Wald's approach is not based on the relation between the geometric equivalents of the properties of matter but is rather based on the {\it signature} of matter {\it on} geometry.

As a second investigation of the spacetime thermodynamics issue we have revisited Jacobson's extraction of Einstein equations by subjecting the approach to Weyl transformations. We saw that in the conformal frame the approach cannot be perceived as fundamental since an observer in such a frame cannot rely on Clausius' thermodynamic relation and use it as it is to extract field equations for spacetime in his/her frame. We saw that only one specific conformal behavior of temperature and entropy --- the latter being at odds with what is implied by Wald's approach --- is allowed to recover the right field equations in the conformal frame. The form of the Clausius relation to be used in the conformal frame does not make any sense when compared to what one expects from such a thermodynamic relation. In addition, we saw that in the presence of a null dust the approach is incapable of reproducing the right field equations in the conformal frame --- as well as in the original frame --- as the source of the field equations in this case can only be recovered up to an arbitrary multiplicative constant times the null dust's energy-momentum tensor.

To track down the root of the issues in extracting Einstein field equations and, hence, of the thermodynamics of spacetime in the conformal frame, we need to recall the fundamental feature of the approach alluded to at the beginning of Sec.~\ref{sec:III}. This consists in assuming from the outset an identity between matter --- or at least properties of matter --- and geometry that only GR is able to correctly justify. In other words, in order to extract Einstein equations one needs, in a sense, to assume these in the first place. This feature is best revealed through Weyl transformations since casting Clausius' relation in terms of geometry can only deform its original material meaning, for geometry behaves differently from matter under such transformations. Within GR, one is able to naturally recover the new field equations after a conformal transformation because one already has a natural clear duality between matter and geometry on both sides of the equations. All one does by performing a conformal transformation on the equations is make manifest on the left-hand side the deformation of spacetime due to the conformal transformation which, when transferred to the right-hand side, appears as an induced source for curvature besides the original material one. Thus, the artificial deformation of spacetime through $\Omega$ is naturally taken into account within the geometric side of the equations. This clear pattern can never be achieved by starting with a deformed Clausius' relation, for the relation originally involves pure material entities on both sides and is ignorant of this crucial dichotomy.

In this paper we have brought to light the deep issues that one faces whenever one tries to understand spacetime from a thermodynamic point of view. This work complements and adds weight to what has been found in Ref.~\cite{AugustPaper}. In the latter reference the investigation was solely centered around the black hole spacetime, whereas in the present work the investigation has been extended to include the more general insights introduced in Refs.~\cite{WaldEntropy,JacobsonPRL} about the thermodynamics of spacetime. As was the case in Ref.~\cite{AugustPaper}, however, our investigation here has only been accomplished at the classical level. These issues might, nevertheless, persist even at the quantum level. Therefore, while a definitive picture can only emerge from a full quantum treatment of spacetime, it is clear that one needs to be careful when reversing the path and attempting to construct such a picture based on the thermodynamics associated to the geometry of spacetime.
\newline
\newline

%%%%%%%%%%%%%%%%%%%%%%%%%%%%%%%%%%%%%%%%%%%%%%%%%%%%%%%%%%%%%%%%%%%%%%%%%%%%%%%%%%%%%%%%%%%%%%%%%%%%%%%%%%%%%%%%%%%%%%%%%%%%%%%%%%%%%%%%%%%%%%%%%%%%%%%%%%%%%%%%%%%%%%%%%%%%%%%%%%%%%%%%%%%%%%%%%%%%%%%%%%%%%%%%%%%%%%%%%%%%%%%%%%%%%%%%%%%%%%%%%%%%%%%%%%%%%%%%%%%%%%%%%%%%%%%%%%%%%%%%%%%%%%%%
\begin{acknowledgments}
The authors are grateful to the anonymous referee for his/her comments that helped improve the clarity of the presentation. This work is supported by the Natural Sciences and Engineering Research
Council of Canada (NSERC) Discovery Grant (RGPIN-2017-05388).
\end{acknowledgments}
%%%%%%%%%%%%%%%%%%%%%%%%%%%%%%%%%%%%%%%%%%%%%%%%%%%%%%%%%%%%%%%%%%%%%%%%%%%%%%%%%%%%%%%%%%%%%%%%%%%%%%%%%%%%%%%%%%%%%%%%%%%%%%%%%%%%%%%%%%%%%%%%%%%%%%%%%%%%%%%%%%%%%%%%%%%%%%%%%%%%%%%%%%%%%%%%%%%%%%%%%%%%%%%%%%%%%%%%%%%%%%%%%%%%%%%%%%%%%%%%%%%%%%%%%%%%%%%%%%%%%%%%%%%%%%%%%%%%%%%%%%
\appendix
%%%%%%%%%%%%%%%%%%%%%%%%%%%%---------------------------------------------------------%%%%%%%%%%%%%%%%%%%%%%%%%%%%%%
\section{Finding $f$ and $W$}\label{A}
In this first appendix we display the detailed calculations that lead one from the fifth line to the last line in Eqs.~(\ref{ConfSteps}). First, we take the divergence of both sides of the fifth line in Eqs.~(\ref{ConfSteps}) and use the contracted Bianchi identity in the conformal frame,  $\tilde{\nabla}^b\tilde{R}_{ab}=\tfrac{1}{2}\tilde{\nabla}_{b}\tilde{R}$, as well as the last equation in the relations (\ref{ConfTransfos}) for the conservation of the conformally transformed energy-momentum tensor: $\tilde{\nabla}^b\tilde{T}_{ab}=-\tilde{T}\tilde{\nabla}_a\Omega/\Omega$. We find,
%%%%%%%%%%%%%%%%%%%%%%%%%%%%--------------------%%%%%%%%%%%%%%%%%%%%%%%%%%%%%%%%%%%%--------------------%%%%%%%%%%%%%%%%%%%%%%%%%%%%%%%%%%%%%%%%%%%%%%%%%%%%%%
\begin{align}\label{DivFifthLine}
&\tfrac{1}{2}\tilde{\nabla}_a\tilde{R}+\tilde{\nabla}_af+\tilde{\nabla}^bW_{ab}= \frac{4\pi\Omega^2}{\eta}\frac{\tilde{\nabla}^b\Omega}{\Omega}\left(\tilde{T}_{ab}-\tfrac{1}{2}\tilde{g}_{ab}\tilde{T}\right)\nonumber\\ 
&= \frac{2\tilde{\nabla}^b\Omega}{\Omega}\left(\tilde{R}_{ab}+f\tilde{g}_{ab}+W_{ab}\right)-\frac{\tilde{\nabla}_a\Omega}{\Omega}\left(\tilde{R}+4f+W\right).
\end{align}
%%%%%%%%%%%%%%%%%%%%%%%%%%%%--------------------%%%%%%%%%%%%%%%%%%%%%%%%%%%%%%%%%%%%--------------------%%%%%%%%%%%%%%%%%%%%%%%%%%%%%%%%%%%%%%%%%%%%%%%%%%%%%
In the second line we have substituted back the expressions of $\tilde{T}_{ab}$ and its trace $\tilde{T}$ as given originally by the fifth line of Eqs.~(\ref{ConfSteps}). We have here denoted by $W$ the trace of the unknown tensor $W_{ab}$. It is now clear from this equation that all the scalar $f$ can include from geometry, i.e., among the derivatives of the metric, is the Ricci scalar $\tilde{R}$ in order to cancel the very first term, for otherwise there would be an extra constraint on geometry besides the final equations of motion. More specifically, the scalar $f$ should be the sum of $-\tfrac{1}{2}\tilde{R}$ plus another scalar that depends only on the conformal factor $\Omega$ and its derivatives. 
\newline
\newline
Let us therefore choose the following Ans\"atze for the scalar $f(\Omega)$ and the 2-tensor $W_{ab}(\Omega)$, respectively: %%%%%%%%%%%%%%%%%%%%%%%%%%-----------------%%%%%%%%%%%%%%%%%%%%%%%%%%%%%%%%%%%%%%%-----------------%%%%%%%%%%%%%%%%%%%%%%%%%%%%%%%%%%%%%%%%%%%%%%%%%%%%%%%%
\begin{align}\label{AnsatzForf&A}
f&=-\tfrac{1}{2}\tilde{R}+\alpha +\beta\,\tilde{\nabla}_{c}\Omega\tilde{\nabla}^c\Omega+\gamma\,\tilde{\Box}\Omega,\nonumber\\
W_{ab}&=\rho\,\tilde{\nabla}_{a}\Omega\tilde{\nabla}_{b}\Omega+\tau\,\tilde{\nabla}_{a}\tilde{\nabla}_{b}\Omega.
\end{align}
%%%%%%%%%%%%%%%%%%%%%%%%%%-----------------%%%%%%%%%%%%%%%%%%%%%%%%%%%%%%%%%%%%%%%-----------------%%%%%%%%%%%%%%%%%%%%%%%%%%%%%%%%%%%%%%%%%%%%%%%%%%%%%%%%
The factors $\alpha$, $\beta$, $\gamma$, $\rho$ and $\tau$ (to be determined) are all scalars that depend {\it a priori} on the metric, the conformal factor $\Omega$ and their derivatives. This specific choice for our Ans\"atze is dictated by the fact that both $f$ and $W_{ab}$ can only depend at most on the second derivatives of the scalar field $\Omega$ in order to guarantee a second-order differential equation of motion for the theory.

Next, with the help of the following two familiar geometric identities:
\begin{align}
    \tilde{\nabla}_a\tilde{\nabla}_b\Omega&=\tilde{\nabla}_b\tilde{\nabla}_a\Omega,\nonumber\\
    \tilde{\nabla}^b\tilde{\nabla}_a\tilde{\nabla}_b\Omega&=\tilde{\nabla}_a\tilde{\Box}\Omega+\tilde{\nabla}^b\Omega\,\tilde{R}_{ab},
\end{align}
and after substituting expressions (\ref{AnsatzForf&A}) for $f$ and $W_{ab}$ into Eq.~(\ref{DivFifthLine}), the latter takes the following form:
\begin{widetext}
\begin{align}\label{AnsatzSubstituted}
&\tilde{\nabla}_a\alpha+\tilde{\nabla}_a\beta\,\tilde{\nabla}_b\Omega\tilde{\nabla}^b\Omega+2\beta\,\tilde{\nabla}_a\tilde{\nabla}_b\Omega\tilde{\nabla}^b\Omega+\tilde{\nabla}_a\gamma\,\widetilde{\Box}\Omega+\gamma\tilde{\nabla}_a\widetilde{\Box}\Omega+\tilde{\nabla}^b\rho\;\tilde{\nabla}_a\Omega\tilde{\nabla}_b\Omega+\rho\;\tilde{\nabla}_a\tilde{\nabla}_b\Omega\tilde{\nabla}^b\Omega+\rho\;\tilde{\nabla}_a\Omega\,\widetilde{\Box}\Omega\nonumber\\
&+\tilde{\nabla}^b\tau\;\tilde{\nabla}_a\tilde{\nabla}_b\Omega+\tau\tilde{\nabla}_a\tilde{\Box}\Omega+\tau\tilde{\nabla}^b\Omega\,\tilde{R}_{ab}=\frac{2\tilde{\nabla}^b\Omega}{\Omega}\;\tilde{R}_{ab}-2\frac{\tilde{\nabla}_a\Omega}{\Omega}\alpha-2\beta\frac{\tilde{\nabla}_a\Omega}{\Omega}\tilde{\nabla}_b\Omega\tilde{\nabla}^b\Omega-2\gamma\frac{\tilde{\nabla}_a\Omega}{\Omega}\tilde{\Box}\Omega+\rho\frac{\tilde{\nabla}^b\Omega}{\Omega}\tilde{\nabla}_a\Omega\tilde{\nabla}_b\Omega\nonumber\\
&+2\tau\frac{\tilde{\nabla}^b\Omega}{\Omega}\tilde{\nabla}_a\tilde{\nabla}_b\Omega-\tau\frac{\tilde{\nabla}_a\Omega}{\Omega}\tilde{\Box}\Omega.
\end{align}
\end{widetext}

In order for this identity to not introduce any extra constraint between the conformal factor $\Omega$ and the metric $\tilde{g}_{ab}$ the identity needs to be satisfied trivially. Therefore, by first comparing the terms that contain $\tilde{R}_{ab}$, we immediately deduce that $\tau=2/\Omega$. Next, by noticing that only two terms, the first on the left and the second on the right, contain a single derivative of $\alpha$ we deduce that $\alpha=\Lambda/\Omega^2$ for some arbitrary constant $\Lambda$. Also, by comparing the terms that have in common the third-order derivative $\tilde{\nabla}_a\tilde{\Box}\Omega$, we deduce that $\gamma+\tau=0$ which implies that $\gamma=-2/\Omega$. Similarly, by comparing the terms that have in common the product $\tilde{\nabla}_a\Omega\,\tilde{\Box}\Omega$, we deduce, based on the values of $\tau$ and $\gamma$ we just found, that $\rho=0$. Finally, by using these values of $\tau$, $\gamma$ and $\rho$ and then comparing the terms that have in common the product $\tilde{\nabla}_a\tilde{\nabla}_b\Omega\,\tilde{\nabla}^b\Omega$, we deduce that $\beta=3/\Omega^2$. 

As a consistency check, we easily verify that the leftover term, $(\tilde{\nabla}_a\beta+2\beta\tilde{\nabla}_a\Omega/\Omega)\tilde{\nabla}_c\Omega\tilde{\nabla}^c\Omega$, vanishes identically after using the value of $\beta$ we just found. Putting all these scalars together inside the Ans\"atze (\ref{AnsatzForf&A}) for $f$ and $W_{ab}$ and substituting these into the fifth line of Eqs.~(\ref{ConfSteps}) gives back indeed the form (\ref{ConfEQ}) of the equations with the right induced energy-momentum tensor $\tilde{T}_{ab}^\Omega$ as given by Eq.~(\ref{InducedEMT}). This completes our derivation. $\Box$
%%%%%%%%%%%%%%%%%%%%%%%%%%%%---------------------------------------------------------%%%%%%%%%%%%%%%%%%%%%%%%%%%%%%
\section{In the presence of a null dust}\label{B}
In the presence of a null dust with energy-momentum tensor $T_{ab}^{\rm null}$, the fifth line in Eqs.~(\ref{ConfSteps}) would read instead,
%%%%%%%%%%%%%%%%%%%%%%%%%%%%--------------------%%%%%%%%%%%%%%%%%%%%%%%%%%%%%%%%%%%%--------------------%%%%%%%%%%%%%%%%%%%%%%%%%%%%%%%%%%%%%%%%%%%%%%%%%%%%%%
\begin{equation}\label{FifthLineNullDust}
\tilde{R}_{ab}+f\tilde{g}_{ab}+W_{ab}= \frac{2\pi}{\eta}\left(\Omega^2\tilde{T}_{ab}+c\tilde{T}_{ab}^{\rm null}\right),
\end{equation}
%%%%%%%%%%%%%%%%%%%%%%%%%%%%--------------------%%%%%%%%%%%%%%%%%%%%%%%%%%%%%%%%%%%%--------------------%%%%%%%%%%%%%%%%%%%%%%%%%%%%%%%%%%%%%%%%%%%%%%%%%%%%%
for any arbitrary constant $c$. Taking the divergence of both sides of this equation, as we did in Appendix \ref{A}, we find, after taking into account that the null dust is traceless and that its energy-momentum tensor is conserved also in the conformal frame, $\tilde{\nabla}^b\tilde{T}_{ab}^{\rm null}=0$,
%%%%%%%%%%%%%%%%%%%%%%%%%%%%--------------------%%%%%%%%%%%%%%%%%%%%%%%%%%%%%%%%%%%%--------------------%%%%%%%%%%%%%%%%%%%%%%%%%%%%%%%%%%%%%%%%%%%%%%%%%%%%%%
\begin{align}\label{FifthLineNullDustBis}
&\tfrac{1}{2}\tilde{\nabla}_a\tilde{R}+\tilde{\nabla}_af+\tilde{\nabla}^bW_{ab}= \frac{4\pi\Omega^2}{\eta}\frac{\tilde{\nabla}^b\Omega}{\Omega}\left(\tilde{T}_{ab}-\tfrac{1}{2}\tilde{g}_{ab}\tilde{T}\right)\nonumber\\ 
&=\frac{2\tilde{\nabla}^b\Omega}{\Omega}\left(\tilde{R}_{ab}+f\tilde{g}_{ab}+W_{ab}+\frac{2\pi c}{\eta}\tilde{T}_{ab}^{\rm null}\right)\nonumber\\
&\quad-\frac{\tilde{\nabla}_a\Omega}{\Omega}\left(\tilde{R}+4f+W\right).
\end{align}
In the second equality we have substituted back the expressions of $\tilde{T}_{ab}$ and its trace $\tilde{T}$ as given by Eq.~(\ref{FifthLineNullDust}). 

Next, recall that in order to guarantee the existence of a Killing vector in the conformal frame, the conformal factor $\Omega$ has to satisfy $\tilde{\xi}^a\tilde{\nabla}_a\Omega=0$ \cite{AugustPaper}. This, in turn, is equivalent to $\tilde{k}^a\tilde{\nabla}_a\Omega=0$. Consequently, we thus also have the following identity involving the null dust's energy-momentum tensor $\tilde{\nabla}^b\Omega\, \tilde{T}_{ab}^{\rm null}=0$. This implies that the fourth term inside the first pair of parentheses in the second equality of Eq.~(\ref{FifthLineNullDust}) cancels. We are thus left with the same equation as Eq.~(\ref{AnsatzSubstituted}), from which the same scalar $f$ and the same 2-tensor $W_{ab}$ as the ones obtained in Appendix \ref{A} also emerge here. 

Substituting these into Eq.~(\ref{FifthLineNullDust}), it follows that the correct induced energy-momentum tensor $\tilde{T}_{ab}^\Omega$ of Eq.~(\ref{ConfEQ}) emerges again in the final field equations but that, as sources of curvature, one obtains, besides the original matter fields on the right-hand side, the null dust's energy-momentum tensor multiplied by an arbitrary constant $c$. This completes our proof of Eq.~(\ref{NullDustIssue}). $\Box$
%%%%%%%%%%%%%%%%%%%%%%%%%%%%--------------------%%%%%%%%%%%%%%%%%%%%%%%%%%%%%%%%%%%%--------------------%%%%%%%%%%%%%%%%%%%%%%%%%%%%%%%%%%%%%%%%%%%%%%%%%%%%%

%%%%%%%%%%%%%%%%%%%%%%%%%%%%%%%%%%%%%%%%%%%%%%%%%%%%%%%%%%%%%%%%%%%%%%%%%%%%%%%%%%%%%%%%%%%%%%%%%%%%%%%%%%%%%%%%%%%%%%%%%%%%%%%%%%%%%%%%%%%%%%%%%%%%%%%%%%%%%%%%%%%%%%%%%%%
%%%%%%%%%%%%%%%%%%%%%%%%%%%%%%%%%%%%%%%%%%%%%%%%%%%%%%%%%%%%%%%%%%%%%%%%%%%%%%%%%%%%%%%%%%%%%%%%%%%%%%%%%%%%%%%%%%%%%%%%%%%%%%%%%%%%%%%%%%%%%%%%%%%%%%%%%%%%%%%%%%%%%%%%%%%%%%%%%%%%%%%%%%%%%%%%%%%%%%%%%

\end{document}